\documentclass[showpacs,showkeys,amsmath,prc,twocolumn,floatfix,aps]{revtex4-1}
\usepackage{graphicx,color}
\usepackage{amssymb}
\usepackage{enumerate}
\usepackage{verbatim}

\usepackage{epstopdf}
\usepackage{amsmath} 
\usepackage{braket}
\usepackage{bm}
\usepackage{multirow}

\begin{document}

\newcommand {\nc} {\newcommand}
\nc {\IR} [1] {\textcolor{red}{#1}}

\title{Transfer reactions and the dispersive optical-model}
\date{\today}
\author{N.~B.~Nguyen$^{1,2}$}
\author{ S.~J.~Waldecker$^{3}$}
\author{F.~M.~Nunes$^{1,2}$}
\author{R.~J.~Charity$^{4}$}
\author{W.~H.~Dickhoff$^{3}$}
 \affiliation{$^1$National Superconducting Cyclotron Laboratory, Michigan State University, East Lansing, Michigan 48824, USA}
 \affiliation{${}^{2}$Department of Physics and Astronomy, Michigan State University, East Lansing, Michigan 48824, USA}
 \affiliation{${}^3$Department of Physics, Washington University, St. Louis, Missouri 63130, USA}
\affiliation{$^{4}$Department of Chemistry, Washington University, St. Louis, Missouri 63130, USA}

\begin{abstract}
The dispersive optical-model is applied to transfer reactions. A systematic study of $(d,p)$ reactions on closed-shell nuclei
using the finite-range adiabatic reaction model is performed at several beam energies and results are compared to data
as well as to predictions using a standard global optical-potential. Overall, we find that the dispersive optical-model is
able to describe the angular distributions as well as or better than the global parameterization. In addition, it also
constrains the overlap function. Spectroscopic factors extracted using the dispersive optical-model are generally
lower than those using standard parameters, exhibit a reduced dependence on beam energy, and are more in line with results obtained from $(e,e'p)$ measurements.
\end{abstract}
\maketitle

\section{Introduction}
\label{introduction}

Rare isotopes have posed new challenges to our understanding of nuclei as  protons and neutrons organized into
well-defined orbitals. Nucleon-nucleon (NN) correlations manifest themselves in different forms as one moves away from the valley of stability.
Nuclear reactions form one of the most important probes for
the limits of stability and the role of the underlying fundamental interactions. Transfer reactions
offer a plethora of opportunities for studying shell structure because one can access ground as well
as excited states (including resonances), and by choosing appropriately the kinematic conditions,
one can explore peripheral properties as well as the nuclear surface.
For this reason, they constitute an important part of the science program of many rare isotopes facilities
worldwide. Examples of recent studies include $^{132}$Sn$(d,p)^{133}$Sn~\cite{sn132dp}, $^{15}$C$(d,p)^{16}$C~\cite{c15dp}, and $^{34,36,46}$Ar$(p,d)$~\cite{ar-exp}.

The increased interest in using transfer reactions as a tool to study
these exotic nuclei has called for reaction-theory developments~\cite{muk05,pang06,deltuva07,moro09,nunes11}.
Whereas in the sixties the standard approach to one-nucleon transfer was to rely on the first-order distorted-wave Born approximation (DWBA),
it is now well understood that in $(d,p)$ reactions, breakup is important and first-order perturbation
is insufficient to describe the process. A practical method for including deuteron breakup was
introduced by Johnson and Soper~\cite{soper,johnson-ria} within a zero-range approximation, usually referred to as
the adiabatic wave approximation (ADWA). In addition to including deuteron breakup to all orders, the ADWA depends exclusively
on nucleon optical-potentials rather than the more ambiguous deuteron optical-potential as in DWBA. ADWA was later extended by
Tandy and Johnson~\cite{tandy} to include finite-range effects (FR-ADWA). A recent systematic study of $(d,p)$ reactions within
FR-ADWA~\cite{nguyen} has shown the importance of finite-range effects in considering deuteron breakup in $(d,p)$ reactions.
The method explored in Refs.~\cite{tandy,nguyen} is based on the truncation of a Weinberg expansion. In Ref.~\cite{nunes11b} a systematic
comparison between $(d,p)$ angular distributions for FR-ADWA and those from the exact full Faddeev solution is performed.
The results from FR-ADWA~\cite{nguyen} are within 10\% of the full solution at forward angle,
for a wide range of beam-energies~\cite{nunes11b}.
While the exact full solution is computationally intensive and expensive,
FR-ADWA calculations are clearly of practical use.

The work in \cite{nunes11b} determines the level of accuracy of FR-ADWA concerning the solution of the three-body $n+p+A$ dynamics in the reaction. For the (d,p) reactions on $^{48}$Ca at 19 MeV and 56 MeV effects are 6\% and 3\% respectively \cite{nunes11b}. With this level of accuracy it is now essential to have good control over the ingredients to the problem,
namely the effective interactions. All applications of FR-ADWA so far have relied on global optical potentials
for the nucleon-target interactions and ``standard single-particle parameters'' for the effective interaction
that determines the final neutron bound state. So far, these two types of input have been disconnected.
Future extraction of nucleon properties in rare isotopes will require presently unavailable knowledge of nucleon-target interactions in the continuum as well as bound-state information.

An attractive feature of the recently implemented dispersive optical-model (DOM)~\cite{charity06,charity07,mueller11}
is that it intrinsically connects the scattering and bound states.
The method, first introduced by Mahaux and Sartor~\cite{mahaux}, and also known as the dynamic polarization potential (DPP),
relies on the dispersion relation between the imaginary and real parts of the nucleon self-energy.
The work of Ref.~\cite{charity06} concentrated on the interaction between protons and Ca isotopes and found
larger correlations near the Fermi surface with increasing nucleon asymmetry.
Neutrons were included in the fits reported in Ref.~\cite{charity07}.
That analysis raised questions concerning the standard parameterization of the imaginary part of
the nucleon potentials as a function of asymmetry $(N-Z)/A$, particularly for neutrons.
More recently, a non-local extension
of the DOM has been introduced that allows for a broader application to calculate observables below the Fermi energy~\cite{dickhoff10}.
Insight from \textit{ab initio} calculations of the nucleon self-energy are expected to provide further guidance on the details of non-locality~\cite{waldecker11}.

In the present work we will use only the local version of the DOM.
Its most recent implementation~\cite{mueller11} yields an accurate representation of a wealth of  data in several locations of the chart of nuclides, including nuclei with $Z=20,28$ and $N=28$ (FIT1), nuclei with $N=50$, with $Z=50$ (FIT2), and finally $Z=82$ (FIT3).
When several isotopes or isotones are included in the fit, the observed nucleon asymmetry dependence of the potentials allows for an extrapolation to nuclei that are more exotic~\cite{mueller11}.
The corresponding predictions can thus be probed with future experiments that may allow further refinements.
This feature of data-driven extrapolations to the respective drip lines and its capacity to link both reaction and structure information make the DOM an excellent framework to study rare isotopes.

In DOM fits~\cite{charity06,charity07,mueller11}, nucleon elastic, total and reaction cross section data,
as well as bound-state properties inferred from $(e,e'p)$ measurements, are all included simultaneously. The aim of the present
work is to explore the application of the DOM to transfer reactions. We include a number of representative cases:
$^{40}$Ca$(d,p)^{41}$Ca, $^{48}$Ca$(d,p)^{49}$Ca,  $^{132}$Sn$(d,p)^{133}$Sn,  and $^{208}$Pb$(d,p)^{209}$Pb.
These reactions are studied at various beam energies for which data are available. We note that apart from the
$n-p$ interaction that binds the deuteron, all effective interactions to the problem in the ADWA framework are provided by the DOM.
We compare the results to those obtained using a global optical-potential (CH89 from Ref.~\cite{ch89}) and standard geometry for the bound state.

In Sec.~\ref{theory} we summarize  the important theoretical concepts pertaining to the ADWA and DOM. Details of the ingredients of the reaction description are presented in Sec.~\ref{details} and the results of the transfer calculations are presented in Sec.~\ref{tran-results}. A more detailed discussion on spectroscopic factors is presented in Sec.~\ref{sf}.
Finally, we summarize and draw our conclusions in Sec.~\ref{conclusion}.


\section{Theory}
\label{theory}

\subsection{Finite-range adiabatic-wave approximation}
\label{theory-adwa}

The adiabatic theory of Refs.~\cite{soper,tandy} for $A(d,p)B$ starts from a three-body model of $n+p+A$.
The deuteron scattering wavefunction in the incident channel is obtained by solving the differential equation:
\begin{eqnarray}
[E+i\epsilon - T_{\bm{r}}- T_{\bm{R}}  - U_{nA} &-& U_{pA} - V_{np} ] \Psi^{(+)}(\bm{r},\bm{R})
\nonumber \\
&=& i \epsilon \phi_{d}(\bm{r})\exp(i \bm{K}_d \cdot \bm{R}),
\label{3b-eq}
\end{eqnarray}
with $\bm{r} =\bm{r}_p - \bm{r}_n$ ($\bm{R}=(\bm{r}_n + \bm{r}_p)/2$) being the relative coordinate (center-of-mass coordinate) of the $n-p$ system. The neutron and proton coordinates, which are taken at the center of mass of the target $A$, are given by $\bm{r}_n$ and $\bm{r}_p$, respectively. We take $U_{nA}(\bm{r}_n)$, $U_{pA}(\bm{r}_p)$, and $V_{np}(\bm{r})$ to be the neutron-target, proton-target, and neutron-proton interactions. In this work, we use the Reid  potential ~\cite{rsc} for $V_{np}$  and DOM potentials for $U_{pA}$ and $U_{nA}$ (see Sect.~\ref{theory-dom}).

In this three-body approach, the solution of Eq.~(\ref{3b-eq}) is inserted into the exact transfer matrix element:
\begin{equation}
T = \bra{\phi_{nA}\chi^{(-)}_{pB}} V_{np}+\Delta_{rem}\ket{\Psi^{(+)}}\ ,
\label{t-eq}
\end{equation}
where $\phi_{nA}$ is the bound state of the neutron-target system, and  $\chi_{pB}^{(-)}$ is a proton scattering distorted wave in the outgoing channel. The remnant operator is $\Delta_{rem}=U_{pA}-U_{pB}$. Contributions from this term are often small except for the lighter
systems.

Johnson and Tandy~\cite{tandy} noted from Eq.~(\ref{t-eq}) that the exact three-body wavefunction $\Psi^{(+)}(\bm{r},\bm{R})$ is only needed within the range of the $V_{np}$ interaction, whenever the remnant contributions are negligible. For this reason, the Weinberg states, which form a complete square integrable basis within the range of $V_{np}$, offer an excellent representation for the three-body scattering wavefunction:
\begin{equation}
\Psi^{(+)}(\bm{r},\bm{R}) = \sum_{i=1}^{\infty} \phi_i(\bm{r}) \chi_i(\bm{R}).
\label{weinberg1}
\end{equation}
Here $\phi_i(\bm{r})$ denote Weinberg states, normalized by $\langle \phi_i |  V_{np} | \phi_j \rangle = - \delta_{ij}$ and satisfying:
\begin{equation}
(T_{\bm{r}} + \alpha_i V_{np}) \phi_i(\bm{r}) = -\epsilon_d \phi_i(\bm{r}),
\label{weinberg2}
\end{equation}
where $\epsilon_d$ is the deuteron binding energy and $\alpha_i$ are the eigenvalues.
Inserting Eq.~(\ref{weinberg1}) into Eq.~(\ref{3b-eq}) one arrives at a non-trivial coupled-channels scattering equation.
However, if only the first term in the Weinberg expansion of Eq.(\ref{weinberg1}) is necessary to describe transfer, then Eq.(\ref{3b-eq}) becomes a single-channel optical-model-type equation:
\begin{equation}
(E + \epsilon_d + i \epsilon - T_R - U_{JT}(R)) \chi^{JT}_1(\bm{R})= i \epsilon N_d  \, \exp(i \bm{K}_d  \cdot \bm{R}),
\label{jt-eq}
\end{equation}
where $N_d$ is the normalization coefficient defined as $N_d=-\langle \phi_1 | V_{np} | \phi_d \rangle$.
The potential $U_{JT}$ in Eq.(\ref{jt-eq}) is calculated by a folding procedure:
\begin{equation}
U_{JT}(R) =  -{\langle \phi_1(\bm{r}) | V_{np} (U_{nA}+U_{pA}) | \phi_1(\bm{r}) \rangle} \,
\label{jtpot-eq}
\end{equation}
where, $\phi_1(\bm{r})$ is the lowest eigenfunction of Eq.~(\ref{weinberg2}), proportional to
the deuteron bound-state wavefunction $\phi_d$.
The expansion of Eq.(\ref{weinberg1}) truncated to the first term, is now inserted into the transfer matrix element:
\begin{equation}
T = \bra{\phi_{nA}\chi^{(-)}_{pB}} V_{np}+\Delta_{rem} \ket{ \phi_d [\chi^{JT}_1(\bm{R})/N_d]}.
\label{tmatrix-fr}
\end{equation}
The validity of the truncation has been tested through comparisons with Faddeev solutions and the results suggest that for most cases of interest, the finite-range ADWA works very well~\cite{nunes11b}.

One important advantage of ADWA over DWBA, is that ADWA depends on nucleon optical-potentials rather than
the deuteron optical-potential, which is far more ambiguous.
ADWA also depends on the nucleon binding potential, as does the DWBA.
Since these nucleon optical and binding potentials are connected by a dispersion relation in the DOM approach, it is interesting to test how the DOM potentials
perform in describing transfer angular distributions for a wide range of energies and targets.
In the next subsection we give a brief summary of the recent developments with DOM nucleon optical-potentials.

\subsection{Dispersive optical-model (DOM)}
\label{theory-dom}

The dispersive optical-model~\cite{mahaux} develops a practical representation of the irreducible nucleon self-energy $\Sigma^*$ from the Green's-function approach to the many-body problem~\cite{dickhoff08}.
The real part of the nucleon self-energy or optical-model potential can be decomposed into an energy-independent non-local part and a dynamic contribution (energy-dependent), that can also be non-local, \textit{i.e.},
\begin{equation}
\text{Re}~\Sigma \left( \bm{r},\bm{r}^{\prime };E\right) =\text{Re}~\Sigma
\left( \bm{r},\bm{r}^{\prime };\varepsilon_{F}\right) +\Delta \mathcal{V}(\bm{r},\bm{r}%
^{\prime };E),  \label{eq:self}
\end{equation}%
where $\varepsilon_F$ is the Fermi energy.
The second term, referred to as the dispersive correction, can be obtained from the imaginary part through the subtracted dispersion relation
\begin{eqnarray}
\lefteqn{\Delta \mathcal{V}(\bm{r},\bm{r}^{\prime };E)=}  \label{eq:sdr} \\
&&+\frac{1}{\pi }\mathcal{P}\int \text{Im}~\Sigma \left( \bm{r},\bm{r}^{\prime
};E^{\prime }\right) \left( \frac{1}{E^{\prime }-E}-\frac{1}{E^{\prime
}-\varepsilon_{F}}\right) dE^{\prime },  \notag
\end{eqnarray}%
where $\mathcal{P}$ stands for the principal value and we note the convention to employ the same sign for the imaginary part of the self-energy above and below the Fermi energy~\cite{mahaux}. By definition in Eq.~(\ref{eq:self}), the dispersive correction is zero at the Fermi energy.
 The dispersive correction
varies rapidly around $E_{F}$ and causes the valence single-particle levels
to be focused towards the Fermi energy.

Following a long tradition~\cite{perey62}, the non-local energy-independent term $\text{Re} \Sigma(\bm{r},\bm{r}';\varepsilon_F)$ is approximated by a local energy-dependent form, generally designated as the Hartree-Fock (HF) potential $\mathcal{V}_{HF}(r,E)$.
While it does describe mean-field properties, it is not strictly a HF contribution, since it involves the complete self-energy at one particular energy.
The energy derivative of $\mathcal{V}_{HF}$ encodes a measure of the non-locality, which is related to the momentum-dependent effective mass
\begin{equation}
\frac{\widetilde{m}\left( r,E\right) }{m}=1-\frac{d\mathcal{V}_{HF}(r,E)}{dE}%
,
\end{equation}%
where $m$ is the nucleon mass.

The local approximation of the HF potential necessitates a scaled imaginary potential given by
\begin{equation}
\mathcal{W} = \frac{\widetilde{m}\left( r,E\right) }{m} \text{Im}\, \Sigma
\end{equation}
and a similarly scaled dispersive correction.
For lack of detailed theoretical input, the imaginary part of the self-energy is also approximated by a local potential in keeping with a long-standing tradition of empirical optical potentials.
We note that the dispersive correction is correspondingly local.
The nucleon self-energy, as typically represented in the DOM, can then be written as
\begin{equation}
\mathcal{U}(r,E) = \mathcal{V}_{HF}(r,E) + \Delta \mathcal{V}(r,E) + i\mathcal{W}(r,E) .
\end{equation}

The Fermi energy is defined as
\begin{eqnarray}
\label{Eq:Fermi}
\varepsilon_{F} &=& \frac{\varepsilon_{F}^{+} + \varepsilon_{F}^{-}}{2}\\
\varepsilon_{F}^{+} & = & M_{A+1} - (M_{A} + m)
\label{Eq:Fermi+} \\
\varepsilon_{F}^{-} & = & M_{A} - (M_{A-1} + m) ,
\label{Eq:Fermi-}
\end{eqnarray}
where $\varepsilon_{F}^{+}$ and $\varepsilon_{F}^{-}$ represent the binding energy for adding or
removing a nucleon, or alternatively, the single-particle energies of the
valence particle and hole states.

Since the current implementation of the DOM includes scattering data up to 200 MeV,
a lowest-order relativistic correction is employed in solving the radial wave equation~\cite{nadasen81}
\begin{equation}
\left[ \frac{d^{2}}{d\rho ^{2}}+\left( 1-\frac{\widetilde{\mathcal{U}}\left(
\rho ,E\right) }{E_{tot}-M-m}-\frac{\ell \left( \ell +1\right) }{\rho ^{2}}%
\right) \right]  u\left( \rho \right) =0  \label{Eq:radial}
\end{equation}%
with $\rho =k\,r$, where $k=\frac{M}{E_{tot}}\sqrt{T\left( T+2m\right) }$, $%
T $ is the laboratory kinetic energy, $E_{tot}$ is the total energy in the
center-of-mass frame, and $M$ is the target mass.
The scaled potential is
\begin{equation}
\widetilde{\mathcal{U}}=\gamma \,\mathcal{U}, \gamma =\frac{2\left(
E_{tot}-M\right) }{E_{tot}-M+m}.  \label{Eq:scaledPot}
\end{equation}%
We denote by $\tilde{\varphi}_{n\ell j}\left( r\right) $ bound-state solutions to the radial
wave equation, where $n$ refers to the corresponding state in the $A\pm1$-system.
The normalized wave functions corrected for non-locality
are then given by
\begin{equation}
\overline{\varphi}_{n\ell j}\left( r\right) =\sqrt{\frac{\widetilde{m}\left(
r,\varepsilon_{n\ell j}\right) }{m}}\tilde{\varphi}_{n\ell j}\left( r\right) ,
\label{eq:overlap}
\end{equation}
where $\varepsilon_{n\ell j}$ is the discrete energy solution to Eq.~(\ref{Eq:radial}).
It should be emphasized that these corrected wave functions correspond to
overlap functions for adding or removing particle from the target
nucleus~\cite{dickhoff10}, \textit{e.g.}
\begin{equation}
\sqrt{S_{n \ell j}} \varphi_{n\ell j}(r) = \bra{\Psi^{A+1}_n}a^\dagger_{r
\ell j}\ket{\Psi^A_0} ,
\label{eq:overlpa}
\end{equation}
where $\varphi_{n \ell j}$ is normalized to one and $S_{n \ell j}$
represents the spectroscopic factor (norm of the overlap function).
The solutions $\overline{\varphi}$ therefore still need to be normalized
by the spectroscopic factor which is obtained in the DOM from:
\begin{equation}
S_{n\ell j}=\int_{0}^{\infty }\overline{\varphi}_{n\ell j}^{2}\left( r\right)
\frac{m}{\overline{m}(r,\varepsilon_{n\ell j})}dr,
\label{eq:SF}
\end{equation}%
where the energy-dependent effective mass determines the reduction of strength from the mean-field value and is given by
\begin{equation}
\frac{\overline{m}\left( r,E\right) }{m}=1-\frac{m}{\widetilde{m}\left(
r,E\right) }\frac{d\Delta \mathcal{V}(r,E)}{dE}.
\end{equation}
The result for the spectroscopic factor in Eq.~(\ref{eq:SF}) was derived in Ref.~\cite{mahaux} and was shown to be an excellent quantitative approximation to the corresponding solution of the Dyson equation that incorporates a non-local HF in Ref.~\cite{dickhoff10}.
We will also include results for the root-mean-square (rms) radius given by
\begin{equation}
R^{rms}_{n\ell j}=\sqrt{\int_{0}^{\infty }\overline{\varphi}_{n\ell j}^{2}\left(
r\right) r^{2}dr}
\end{equation}%
in the DOM analysis.

In the present work, we will extract the normalization of the overlap functions from the analysis of the transfer reactions by scaling the theoretical cross section to match the data.
We will also compare these results with the DOM spectroscopic factors generated by Eq.~(\ref{eq:SF}).
For comparison purposes, we will also consider the single-particle wave function produced
with a Woods Saxon potential with standard parameters (radius $r_0=1.25$ fm and diffuseness $a=0.65$ fm) and the depth adjusted to reproduce the experimental binding energies. These wave functions will be referred to as $\varphi_{WS}$.
An important additional experimental constraint on the overlap function at large distances (when available) is provided by an asymptotic normalization constant (ANC). The single particle ANC 
$b_{n \ell j}$ is defined through the asymptotic behavior of the single particle state $\varphi$
\begin{equation}
\varphi_{n \ell j}(r) \Rightarrow b_{n \ell j} i \kappa h_{\ell}(i\kappa r) ,
\label{eq:ANC}
\end{equation}
where $\kappa =\sqrt{2\mu \varepsilon_{n\ell j}}$, $\mu$ is the corresponding reduced mass
and $\varphi_{n \ell j}(r)$ is mormalized to one.
We note that the spherical Hankel function in Eq.~(\ref{eq:ANC}) is only appropriate for neutrons.


\section{Results}
\label{results}

\subsection{Details of the calculations}
\label{details}

\begin{table}
\caption{Properties of overlap functions with a comparison of the DOM overlap function,
corrected for non-locality $\bar{\varphi}$ with the Woods-Saxon single-particle wavefunction $\varphi_{WS}$. These include the counting number $n$, the angular momenta ($\ell j$) of the valence orbital, the separation energy $S_n$, the root mean square radius of the valence orbital $R^{rms}$, and the modulus of the single-particle ANC $b_{nlj}$.
} \label{spprop}
\begin{center}
\begin{tabular}{clcccc}
\hline \hline
Nucleus & Overlap &  $n\ell j$ & $S_n$ [MeV] & $R^{rms}  [fm] $ & $|b_{nlj}|$ [fm$^{-1/2}]$\\ \hline
\multirow{2}{*}{$^{41}$Ca} & $\varphi_{_{WS}}$ & \multirow{2}{*}{$0f_{7/2}$} & \multirow{2}{*}{8.362} & 3.985 & 2.285\\
& $\bar{\varphi}$ & & & 3.965 & 2.261 \\ \hline
\multirow{2}{*}{$^{49}$Ca} & $\varphi_{_{WS}}$ & \multirow{2}{*}{$1p_{3/2}$} & \multirow{2}{*}{5.146} & 4.606 & 5.818\\
& $\bar{\varphi}$ & & & 4.820 & 6.098 \\ \hline
\multirow{2}{*}{$^{133}$Sn} & $\varphi_{_{WS}}$ & \multirow{2}{*}{$1f_{7/2}$} & \multirow{2}{*}{2.469} & 6.080 & 0.844\\
& $\bar{\varphi}$ & & & 6.513 & 0.831 \\ \hline
\multirow{2}{*}{$^{209}$Pb} & $\varphi_{_{WS}}$ & \multirow{2}{*}{$1g_{9/2}$} & \multirow{2}{*}{3.936} & 6.498 & 1.650\\
& $\bar{\varphi}$ & & & 6.746 & 1.827 \\ \hline \hline
\end{tabular}
\end{center}
\end{table}

Separate DOM fits were produced for different parts of the chart of nuclides in Ref.~\cite{mueller11}.
For the Ca isotopes, the DOM fit included nucleon scattering and bound-state data for nuclei with $Z=20,28$ and $N=28$ (FIT1). Excluded were proton elastic scattering data on $^{40}$Ca below 18 MeV, since for those cases the optical model does poorly and there are fluctuations in the reaction cross section~\cite{dicello71}.
For the Sn isotopes, the DOM fit includes proton elastic-scattering data on $^{112-124}$Sn and neutron elastic-scattering data on $^{116}$Sn, $^{118}$Sn, $^{120}$Sn, and $^{124}$Sn (FIT2)~\cite{mueller11}. Finally, for $Z=82$,  $^{208}$Pb data are included (FIT3).
The fitting procedure follows previous works~\cite{charity06,charity07}
but the depth of the HF potential was adjusted to reproduce the valence particle levels for neutrons.
Since transfer cross sections
are strongly dependent on the binding energy of the final bound state, particular effort was made in all these fits, to reproduce
the experimental binding energies. The DOM parameterizations are subsequently used to calculate $A(d,p)B$  within FR-ADWA.
The DOM thus provides all optical potentials $U_{nA}$, $U_{pA}$, $U_{pB}$ as well as the neutron bound-state interaction $V_{nA}$.
The properties of the neutron states considered in this work are summarized in Table~\ref{spprop}, including the rms radius and the single-particle asymptotic normalization coefficient~\cite{muk05} [see Eq.~(\ref{eq:ANC})]. We use {\sc twofnr} \cite{twofnr} to calculate the finite-range deuteron adiabatic potential and {\sc fresco} \cite{fresco} to calculate the transfer cross sections.

\begin{figure}
\includegraphics[width=0.4\textwidth]{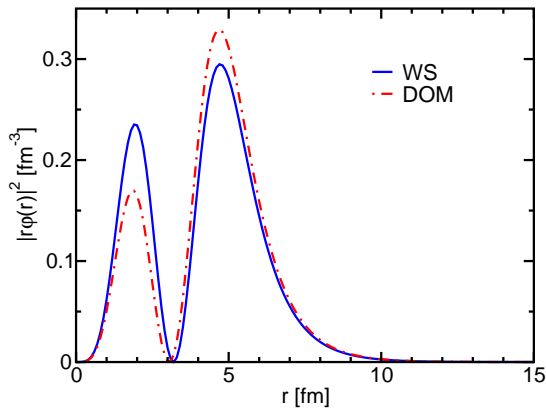}
\caption{(Color online) Comparison of the square of the single-neutron overlap functions for $^{49}$Ca obtained with a Woods-Saxon potential $\varphi_{WS}$ (solid) and with the local DOM corrected for non-locality $\bar{\varphi}$ (dashed). }
\label{fig-overlap}
\end{figure}

The DOM fit for Ca isotopes generated a radius parameter of 1.18 fm while the corresponding standard Woods-Saxon potential used for comparison was fixed at 1.25 fm.
The non-locality correction almost completely cancels this difference making the DOM overlap function $\bar{\varphi}$ for $^{41}$Ca essentially identical to $\varphi_{WS}$.
When a node is present, the non-locality correction is even more pronounced making the rms radius of the $1p_{3/2}$ DOM wave function larger  than its Woods-Saxon counterpart.
We illustrate this in Fig.\ref{fig-overlap} for the $^{49}$Ca ground state, where the square of the single-neutron overlap function
obtained with the local DOM, corrected for non-locality, is compared to the Woods-Saxon counterpart with standard geometry.
The DOM result (FIT2) for the radius parameter for Sn nuclei is 1.24 fm and therefore a substantially larger rms radius is obtained for the DOM overlap function compared to the standard Woods-Saxon one when the non-locality correction is applied for the $1f_{7/2}$ orbit (see Table~\ref{spprop}).

As is usual in optical-model potentials, the imaginary potential in the DOM is separated into surface and volume terms. The former is stronger at the lower energies considered in this work and the latter becomes dominant only at much higher energies.
The volume term was assumed to have a small dependence on $\pm(N-Z)/A$, where the positive and negative signs refer to protons and neutrons.
The magnitude of this term was determined from the fit to the lead data (FIT3 of Ref.~\cite{mueller11}) and then was kept fixed when fitting the Sn and Ca  data~\cite{mueller11}.
It is for the study of $n-^{132}$Sn (derived from FIT2 in Ref.~\cite{mueller11}) that the asymmetry dependence becomes critical, as one is extrapolating well outside the region of fitted data.
As previously seen for the Ca isotopes~\cite{charity07}, and for FIT1 in Ref.~\cite{mueller11}, the Sn DOM potentials contain an insignificant asymmetry dependence of the neutron imaginary surface term, while that for protons exhibited a linear dependence on $(N-Z)/A$.

\begin{figure}[h!]
\centering{\resizebox*{0.4\textwidth}{!}{\includegraphics{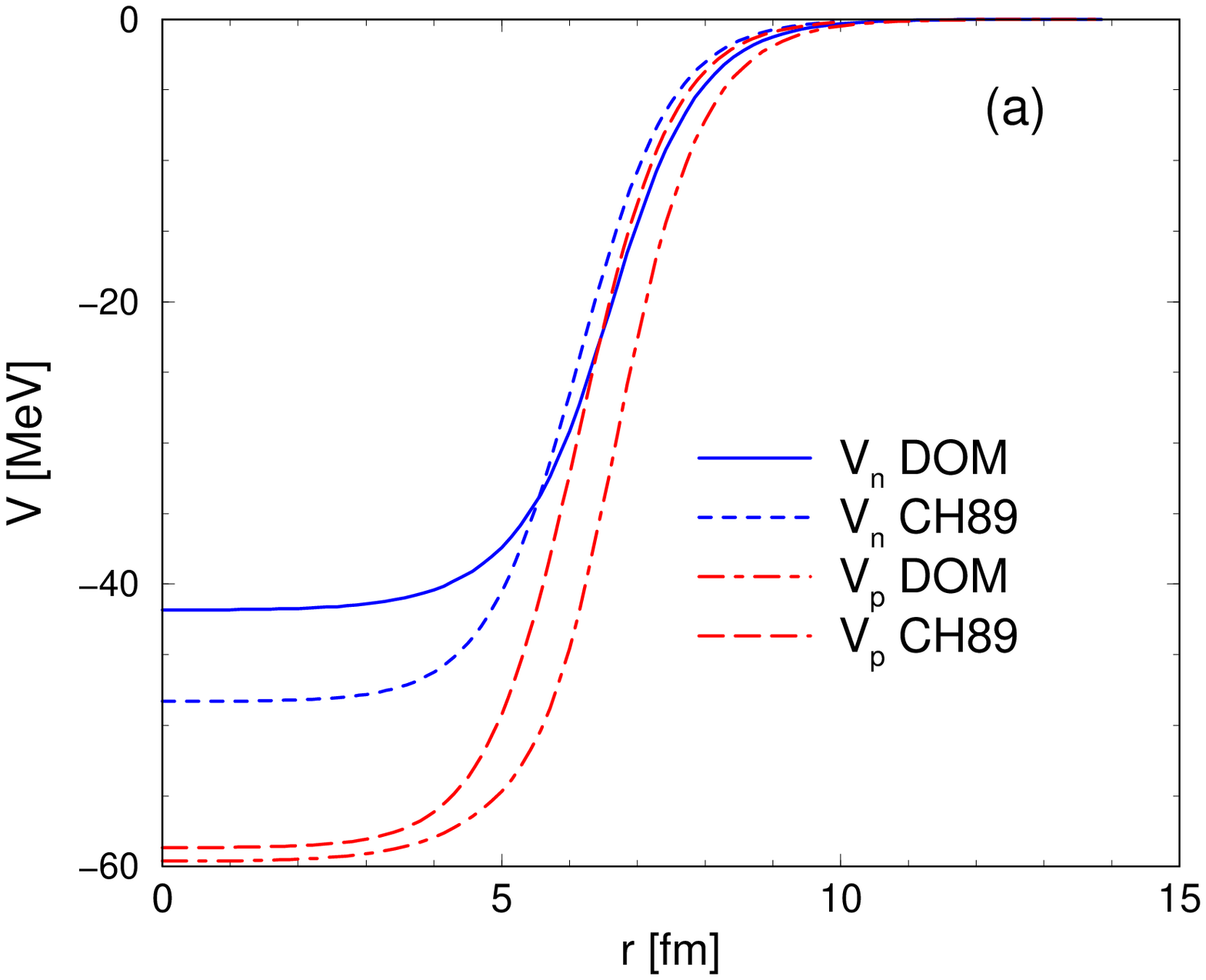}}} \\
\centering{\resizebox*{0.4\textwidth}{!}{\includegraphics{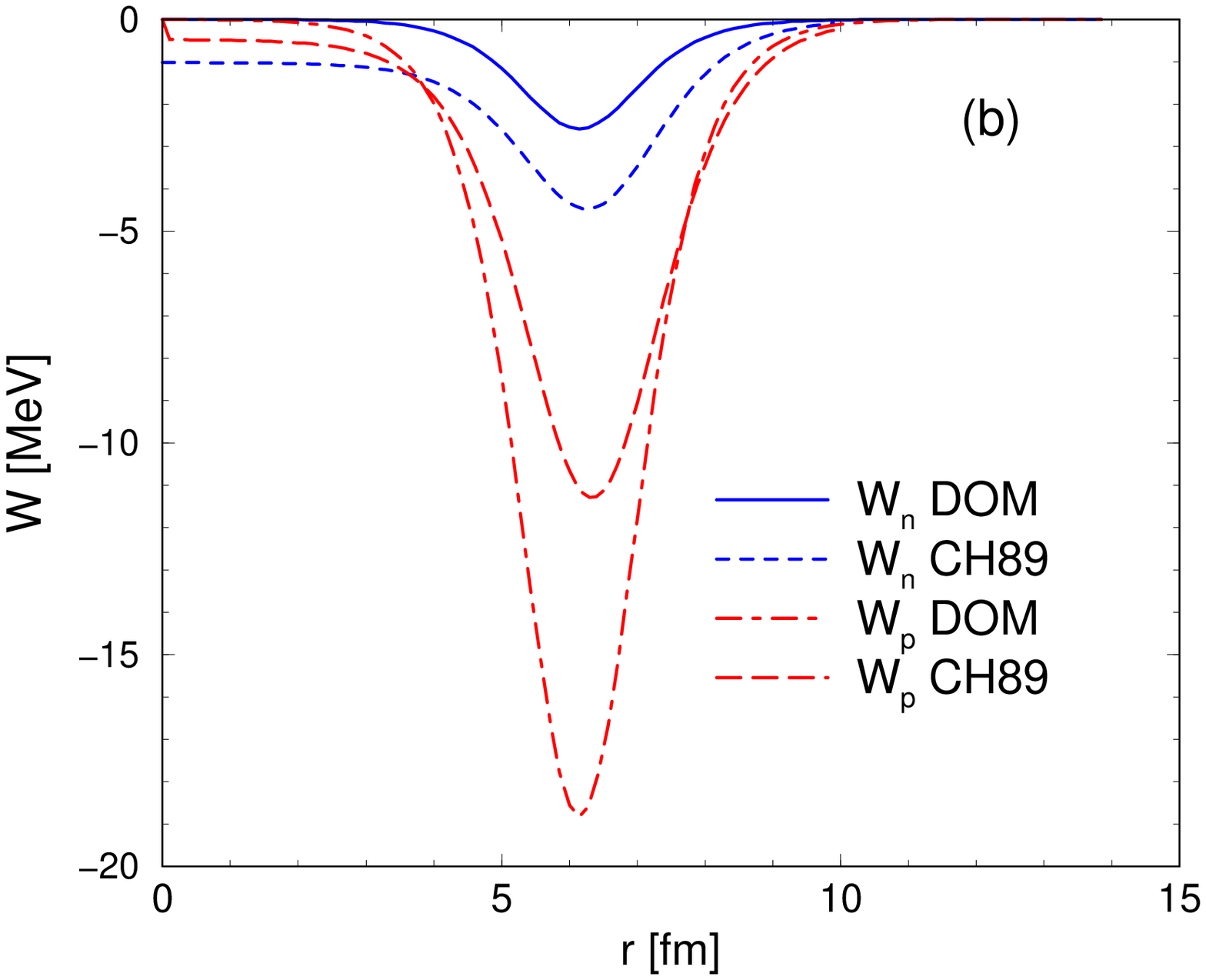}}} \\
\caption{\label{fig-pot}(Color online) Optical potentials for n-$^{132}$Sn
and p-$^{132}$Sn (thin) at $4.7$ MeV: comparing DOM with CH89. 
The real component is shown in panel (a) and the imaginary part in panel (b).}
\end{figure}
A comparison between the DOM and CH89 optical potential reveals some similarities and some differences.
Since the surface absorption for neutrons exhibits a different (weaker) asymmetry dependence than assumed in Ref.~\cite{ch89},
it is not surprising that the ${}^{132}$Sn potentials show the largest differences,
as this corresponds to a substantial extrapolation from potentials constrained by data (see Fig.\ref{fig-pot}).
In particular, the neutron (proton) surface absorption in the DOM is substantially less (more) than in the CH89 parameterization).
The real parts of the CH89 potentials have the same radius parameter for both protons and neutrons by decree, whereas in the DOM, due to the different surface absorption, the dispersive correction makes the real proton potentials extend farther than those for neutrons. 
We also note that the CH89 potentials are not dispersive, making a more detailed comparison less productive.

\subsection{Transfer cross sections}
\label{tran-results}

We consider 9 reactions for which data exists:
$^{40}$Ca$(d,p)^{41}$Ca at $E_d=20$ and $56$ MeV, $^{48}$Ca$(d,p)^{49}$Ca at $E_d=2, 13, 19.3$ and $56$ MeV, $^{132}$Sn$(d,p)^{133}$Sn
at $E_d=9.46$ MeV,  and $^{208}$Pb$(d,p)^{209}$Pb at $E_d=8$ and $20$ MeV.
\begin{figure}[t]
\centering{\resizebox*{0.4\textwidth}{!}{\includegraphics{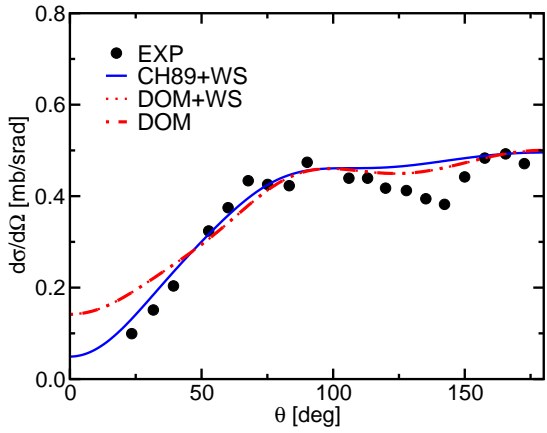}}} \\
\caption{\label{fig-ca48dp1}(Color online) Angular distributions are shown for the reaction $^{48}$Ca$(d,p)^{49}$Ca at $E_d = 2$ MeV.
Theory predictions have been normalized to the data at backward angles.}
\end{figure}

\begin{figure}[h!]
\centering{\resizebox*{0.4\textwidth}{!}{\includegraphics{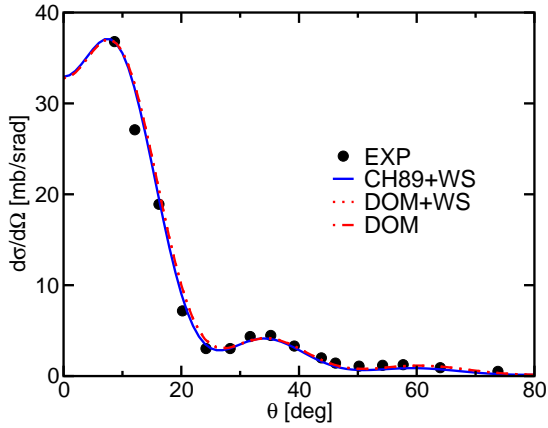}}} \\
\caption{\label{fig-ca48dp2}(Color online) Angular distributions are displayed for the reaction $^{48}$Ca$(d,p)^{49}$Ca at $E_d = 19.3$ MeV.
Theory predictions have been normalized to the data at the peak.}
\end{figure}

\begin{figure}[h!]
\centering{\resizebox*{0.4\textwidth}{!}{\includegraphics{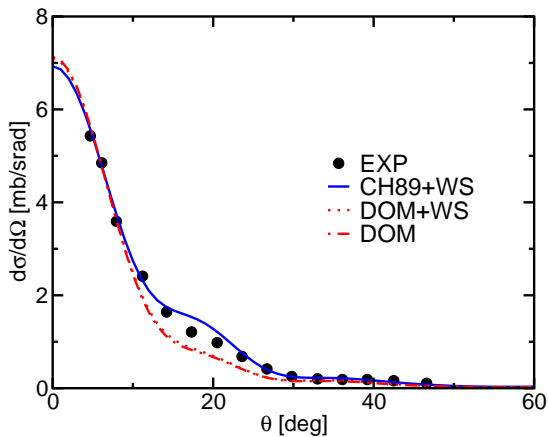}}} \\
\caption{\label{fig-ca48dp3} (Color online) Angular distributions for the reaction $^{48}$Ca$(d,p)^{49}$Ca at
$E_d = 56$ MeV are displayed.
Theory predictions have been normalized to the data at forward angles.}
\end{figure}
FR-ADWA calculations for all these cases were performed for three interaction models: optical potentials from CH89~\cite{ch89} and the neutron
overlap function as $\varphi_{WS}$ (CH89+WS); optical potentials from DOM but the neutron overlap function as  $\varphi_{WS}$ (DOM+WS);
and finally both the optical potentials and the neutron overlap function from this local approximation of DOM corrected
for non-locality $\bar{\varphi}$ (DOM). In Figs.~\ref{fig-ca48dp1}, \ref{fig-ca48dp2}, and~\ref{fig-ca48dp3} the angular distributions for $E_d=2, 19.3$ and $56$ MeV, normalized to data, are plotted for the $^{48}$Ca$(d,p)^{49}$Ca reaction. 

The normalization is performed at the peak of  the data, \textit{i.e.} at backward angles for reactions below the Coulomb barrier
and at the first peak when the bombarding energy is above it.
For these cases the relevant angles were $\theta \approx 155^\circ$, $\theta \approx 10^\circ$ and $\theta \approx 5^\circ$ for $E_d=2, 19.3$ and $56$ MeV, respectively.
Note that at 2 MeV, compound contributions to the transfer cross section can be important. However for this case, a close analysis of the experimental angular distribution suggests such a contribution to be negligible.

\begin{figure}[t!]
\centering
\includegraphics[width=0.4\textwidth]{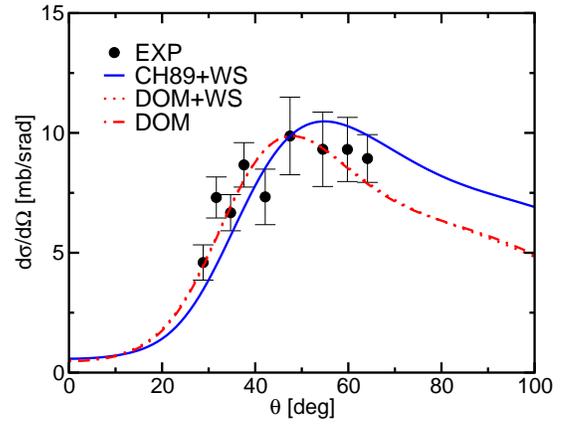}
\caption{(Color online) Angular distributions for the $^{132}$Sn$(d,p)^{133}$Sn reaction at a deuteron energy of $E_d = 9.46$ MeV are shown normalized at the peak of the experimental cross section. }
\label{fig-sn132}
\end{figure}

\begin{figure}[t!]
\centering
\includegraphics[width=0.4\textwidth]{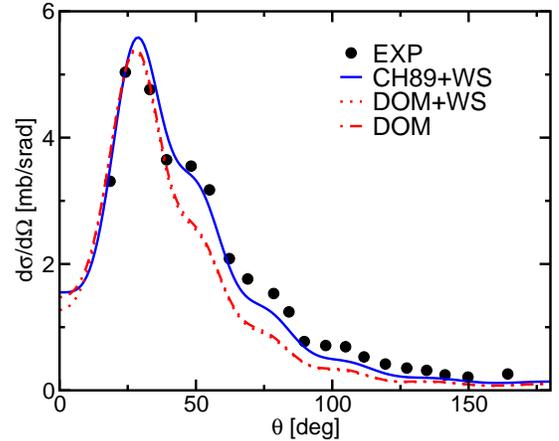}
\caption{(Color online) Angular distributions for the $^{208}$Pb$(d,p)^{209}$Pb reaction are shown at a deuteron energy of $E_d = 20$ MeV and normalized at the peak of the experimental data. }
\label{fig-pb208}
\end{figure}
The first thing to note is that DOM is able to describe the cross sections well. However, there is no significant difference between the angular dependence predicted by DOM, DOM+WS and CH89+WS, even though there are differences in the shapes of the overlap functions as shown in Fig.~\ref{fig-overlap}.
As a similar set of data was used to constrain the CH89 optical potential \cite{ch89},
one can conclude that this set is sufficient to produce the correct angular distribution for $(d,p)$ reactions within FR-ADWA for this nucleus.
For the $(d,p)$ reactions with $^{40}$Ca at the $E_d=20$ and $56$ MeV, DOM performs as well as CH89+WS and no significant difference
is observed between the standard and the dispersive approach (not shown here).

There were two cases for which we did see considerable difference in
the angular distributions between DOM and CH89, the $^{132}$Sn and $^{208}$Pb.
These distributions are shown in Figs.~\ref{fig-sn132} and \ref{fig-pb208}.
It is clear that this is not due to the modification in the shape of the overlap functions, but rather the optical potentials.
The real part of the optical potential in the DOM has a larger radius, which shifts the diffraction
pattern toward smaller angles. However an increase in radius in the overlap
function only affects the magnitude at forward angles (shown by comparing DOM+WS and DOM).
While for $^{132}$Sn DOM appears to improve the angular distribution, for $^{208}$Pb the angular distributions using DOM
optical potentials move away from the data. We will address the reactions with $^{208}$Pb in more detail later.

\begin{table}[b]
\caption{Spectroscopic factors obtained from the FR-ADWA analysis. The deuteron kinetic energy $E_d$ (lab. frame) is in MeV. Reference to the experimental data set used in the extraction is also given \label{sfact}}
\begin{center}
\begin{tabular}{lcc|cccr}
\hline
\hline
Nucleus &                   E$_d$ & data & CH89+WS & DOM+WS & DOM &       DOM(th)\\ \hline
\multirow{2}{*}{$^{41}$Ca} & 20 &  \cite{ca40-20} &  0.96 &   0.85 &  0.86 & \multirow{2}{*}{0.75}\\
                            & 56 & \cite{ca40-56} &   0.88 &   0.73 &  0.74 &\\ \hline
\multirow{4}{*}{$^{49}$Ca} & 2 & \cite{ca48-2} &     0.94 &  0.72 &  0.66 & \multirow{4}{*}{0.80}\\
                            & 13 & \cite{ca48-13} &  0.82 &    0.67 &  0.61 & \\
                            & 19.3 & \cite{ca48-19} & 0.77 &  0.68 &  0.62 & \\
                            & 56 & \cite{ca48-56} &  1.1 &     0.70 &  0.62 & \\ \hline
$^{133}$Sn                  & 9.46 & \cite{sn132dp} & 1.1 &   1.0 &   0.72 &                  0.80\\ \hline
\multirow{2}{*}{$^{209}$Pb} & 8 & \cite{pb208-8} &  1.7 &     1.5 &   1.2 & \multirow{2}{*}{0.76}\\
                            & 20 & \cite{pb208-20} &  0.89 &    0.61 &  0.51 &\\
\hline \hline
\end{tabular}
\end{center}
\end{table}

\subsection{Spectroscopic factors}
\label{sf}

While the angular distributions predicted using DOM do not differ considerably from those using CH89, the normalization
of the cross sections do. We determine an experimental spectroscopic factor by taking the ratio of $d\sigma/d\Omega$(exp) over $d\sigma/d\Omega$(theory) for $\theta$ at the first peak of the distribution for all but sub-barrier energies. At such
energies the normalization is determined at backward angles.
We compare the results obtained in the various approaches in Table~\ref{sfact}: the labels CH89+WS, DOM+WS and DOM
are exactly as described in Sec.~\ref{tran-results}. Note that the normalizations are calculated with an overlap function
normalized to unity. However, the DOM predicts overlap functions that are not normalized to unity, as correlations are already taken into account.
The spectroscopic factor coming directly from such an analysis is given in the last column of
Table~\ref{sfact} and labeled DOM(th).

One should keep in mind that for $^{40}$Ca, it is notoriously difficult to describe low-energy scattering. This fact led to
the exclusion of proton elastic scattering data at energies below 18 MeV from the DOM fit, as discussed above.
Since in the ADWA the optical potentials are evaluated at $E_d / 2$, the DOM results for $^{40}$Ca$(d,p)$ at $E_d = 20$ MeV are not well constrained by elastic nucleon scattering data.
In all other cases, DOM potentials provide a good description of scattering observables.

If one focuses first on the traditional CH89+WS approach, one sees that a wide range of spectroscopic factors can be obtained depending
on the beam energy. For example with ${}^{48}$Ca, the spectroscopic factor ranges from $S=0.77$ to 1.1.
This unwanted energy dependence was already seen in the systematic study in \cite{lee07}. 
Note that spectroscopic factors extracted in \cite{lee07} do not coincide with those in our Table~\ref{sfact} due to additional approximations in the reaction theory in \cite{lee07}, namely the zero-range ADWA corrected within the local energy approximation (errors introduced in making those approximations have been quantified in \cite{nguyen}).
The large energy dependence is significantly reduced when DOM optical potentials are used,
the exception being the the reactions on $^{208}$Pb, which we will consider later.

When comparing CH89+WS and DOM+WS, one sees that the DOM optical potentials reduce the spectroscopic factor by a non-negligible amount.
Note that this is due to the optical potentials alone, since the overlap functions are the same in CH89+WS and DOM+WS.
If one now replaces the single-particle wave function by that predicted from DOM, two things can happen: virtually no change
(as in the case of $^{40}$Ca) or an important further reduction of the spectroscopic factor (as in all other cases).
It is indeed only for ${}^{40}$Ca that the wave functions are similar.
In all other cases, the different radius parameter obtained in DOM fits, together with the non-locality correction, shift density from the interior and enhance the probability in the
surface region.
A larger overlap function at the surface, produces larger cross sections which then
imply smaller spectroscopic factors.
Values extracted using DOM ingredients are much more in line with those from $(e,e'p)$ measurements~\cite{lapikas93} with the exception of $^{208}$Pb.

Finally, we also show the spectroscopic factors predicted directly by the DOM without reference to transfer data (last column).
These are larger than those extracted from experiment for ${}^{48}$Ca and ${}^{132}$Sn.
This difference is mostly associated with the different choice in Ref.~\cite{mueller11} for the domain where the imaginary part of the optical potentials vanish as compared to the choice in Ref.~\cite{charity07}.
This leads to a larger particle-hole gap in Ref.~\cite{mueller11} as compared to Ref.~\cite{charity07} for the same nuclei.
Since spectroscopic factors are particularly sensitive to the particle-hole gap, an increase of the order of 0.1 is obtained in Ref.~\cite{mueller11} somewhat similar to the difference between columns DOM and DOM(th) in Table~\ref{sfact} for these nuclei.
If one adopts the perspective that for nuclei where optical potentials relevant for transfer reactions are well-constrained and the FR-ADWA provides an accurate description, the present results suggest that future DOM fits should either reduce the particle-hole gap or make the coupling at low energy stronger.
Nevertheless, the consistency of the extracted spectroscopic factors employing DOM ingredients is very encouraging (see also the discussion for ${}^{208}$Pb below).

\begin{table}
\caption{ The square of the many-body asymptotic normalization coefficient $C^2$ for various
models (fm$^{-1}$).} \label{anc}
\begin{tabular}{lccccr}
\hline \hline
Nucleus & $E_d$ [MeV] & CH89+WS & DOM+WS & DOM & DOM(th)\\ \hline
 \multirow{2}{*}{$^{41}$Ca} & 20 & 5.0 & 4.4 & 4.4 &\multirow{2}{*}{2.8}\\
 & 56 & 4.6 & 3.8 & 3.8 &\\ \hline
\multirow{4}{*}{$^{49}$Ca} & 2 & 31.7 & 24.4 & 24.4 & \multirow{2}{*}{29.6} \\
& 13 & 27.9 & 22.7 & 22.6 &\\
& 19.3 & 26.0 & 23.1 & 23.0 &\\
& 56 & 35.8 & 23.5 & 23.2 &\\ \hline
$^{133}$Sn & 9.46 & 0.78 & 0.71 & 0.49 & 0.56 \\ \hline
\multirow{2}{*}{$^{209}$Pb} & 8 & 4.5 & 4.1 & 4.2 &\multirow{2}{*}{2.5}\\
& 20 & 2.4 & 1.7 & 1.7 &\\
\hline \hline
\end{tabular}
\end{table}
Most of the reactions studied in this work are not sensitive to details of the overlap function in the interior,
but rather to the ANC~\cite{pang06}. For completeness, in Table \ref{anc} we present
the square of the many-body ANCs obtained directly from the spectroscopic factors
presented in Table \ref{sfact}, multiplied by the square of the single-particle ANC ($C^2=S b^2$).
The ANC for $^{41}$Ca was determined to be
$C_{f7/2}^2=8.36 \pm 0.42$ fm$^{-1}$, using a DWBA analysis of sub-Coulomb $(d,p)$ data~\cite{pang06}. 
The analysis in Ref.~\cite{pang06} shows that with that large
ANC, there is no consistency between the expected spectroscopic factor for $^{41}$Ca and the $^{40}$Ca$(d,p)^{41}$Ca data
above the Coulomb barrier. In Ref.~\cite{pang06} it is also shown that to obtain consistency between sub-Coulomb
and above Coulomb energies one needs an ANC about half that value. This work solves the discrepancy seen for $^{41}$Ca in Ref.~\cite{pang06}.
The ANC for $^{49}$Ca was determined in Ref.~\cite{muk08} from a DWBA analysis of sub-Coulomb $(d,p)$ data to be $C_{p3/2}^2=32.1 \pm 3.2$ fm$^{-1}$.
This value is consistent with the values we obtain using CH89+WS. 
Using the DOM instead, the ANC in Ref.~\cite{muk08} should be significantly
reduced. An analysis of the $^{132}$Sn$(d,p)^{133}$Sn reaction around the Coulomb barrier~\cite{jones11} using FR-ADWA and the CH89 optical
potentials generates an ANC for the ground state of $C_{f7/2}^2=0.82 \pm 0.07$ fm$^{-1}$. This value would significantly be reduced if DOM were used instead. Finally, in Ref.~\cite{muk05} an ANC for $^{209}$Pb is extracted from heavy ion reactions $C_{g9/2}=2.15 \pm 0.16$ fm$^{-1}$
using DWBA. This value is close to the DOM theoretical prediction.

The transfer results for $^{208}$Pb are problematic for two reasons: i) there is a very large discrepancy between the spectroscopic factors obtained at $E_d=20$ MeV and $E_d=8$ MeV, and ii) for the sub-Coulomb barrier energy, this value 
is much larger than unity.
No issues arose in the DOM fitting for this nucleus and therefore the DOM fits can be considered reliable for the elastic and total cross-section data. Even though the transfer angular distributions are well described within FR-ADWA, we cannot exclude the possibility of
target excitation. In order to test whether the cause for inconsistency is the reaction mechanism, we performed exploratory calculations including the low-lying $3^-$ and $2^+$ states in $^{208}$Pb and the strong octupole and quadrupole couplings between these states and the ground state within a coupled-channel Born approximation (CCBA), using a global optical potential for the deuteron \cite{lh}. One should keep in mind that,
within a CCBA approach, the extraction of the spectroscopic factor for a single orbital is obscured by other components.
We found the effects to be strong for $^{208}$Pb(d,p) at 20 MeV but rather weak at the lower energy.
We also performed coupled reaction channel calculations, iterating the transfer coupling, but
again found the effects to be weak at the lower energy while significant at higher energy.
These results suggest that at least for the higher energy, the reaction mechanism goes
beyond the present implementation of ADWA. The unrealistic SF obtained at the lower energy could indicate the failure
of ADWA. At present, for reactions involving such large Coulomb fields, 
we cannot verify the validity of ADWA through the comparison with exact Faddeev calculations. The present techniques
used in solving the Faddeev equations are limited to targets with $Z\approx 20$.

If the reaction mechanism for $^{208}$Pb$(d,p)$ at 20 MeV necessitates the inclusion of target excitation, could this mechanism also play a role in our
other test cases, even if the angular distributions are well described in our present formulation?
In Ref.~\cite{pang06} a study of target excitation in $^{40}$Ca$(d,p)^{41}$Ca at low energy shows strong effects.
However we expect these to be small at $56$ MeV. The $^{48}$Ca nucleus has a very weak transition to its first excited
state, and therefore no significant effect of target excitation is expected. Target excitation was tested for
the reaction on $^{132}$Sn and found to be negligible.
Nevertheless, the present results for $^{208}$Pb do call for an extension of the FR-ADWA to include target excitation
and deuteron breakup in a consistent framework.


\section{Conclusions}
\label{conclusion}
We have tested the performance of the dispersive optical-model potentials and corresponding overlap functions in $(d,p)$ reactions.
We performed finite-range adiabatic calculations for a range of closed-shell nuclei covering a wide range of beam energies.
Within this description, the DOM provides all necessary ingredients apart from the deuteron $V_{np}$ interaction,
\textit{i.e.} nucleon optical-potentials and the mean-field binding the neutron in the final state.
We compare the results obtained with the DOM to those obtained using a global parameterization of the nucleon optical-potential and a standard geometry for the neutron single-particle overlap.
We find that the DOM performs as well as the CH89 parameterization in the description of the angular distributions.
While spectroscopic factors extracted within the standard approach can be strongly dependent
on the energy at which the $(d,p)$ data were obtained, this dependence is strongly suppressed when DOM potentials are employed.
The exception is $^{208}$Pb, a case for which coupled-channel calculations
demonstrate the importance of target excitation.
Overall, we find that the extracted spectroscopic factors using DOM are significantly reduced when compared to the standard
approach, bringing the values more in line with those obtained from $(e,e'p)$ measurements.

Because DOM potentials can be extrapolated to rare isotopes and separate checks of their quality can be made by performing elastic nucleon (proton) scattering experiments in inverse kinematics, the present framework of the DOM in which reaction and structure data are incorporated on the same footing, provides an excellent platform to analyze transfer reactions involving rare isotopes.
\acknowledgments The work of N.B.N. and F.M.N. was partially supported by the National Science Foundation grant PHY-0555893, the Department of Energy through grant DE-FG52-08NA28552
and the TORUS collaboration DE-SC0004087. The work of S.J.W. and W.H.D. was supported by the National Science Foundation under grant PHY-096894, and R.J.C by the U.S. Department of Energy, Division of Nuclear Physics under grant FG02-87ER-40316.

\end{document}